\documentclass[reqno,12pt]{article} 
\usepackage[margin=2cm]{geometry}   
\usepackage{amsmath,amsthm,amssymb,amscd,amstext,amsfonts}
\usepackage{mathtools}
\usepackage{mathrsfs}
\usepackage{thmtools}
\usepackage{latexsym}
\usepackage{verbatim}
\usepackage{framed}
\usepackage{graphicx}
\usepackage{stmaryrd}
\usepackage{enumerate}
\usepackage{fullpage}
\usepackage{bm}
\usepackage{color}
\usepackage{hyperref}
\usepackage{url}
\usepackage{physics}

\newtheorem{theorem}{Theorem}

\newtheorem{definition}{Definition}

\theoremstyle{remark}

\DeclareMathOperator{\Ex}{\mathbb{E}}           

\DeclareMathOperator{\AND}{AND}
\DeclareMathOperator{\OR}{OR}

\DeclareMathOperator{\bs}{bs}        

\newcommand{\N}{\mathbb{N}}

\newcommand{\cA}{\mathcal A}

\newcommand{\cT}{\mathcal T}

\begin{document}

\title{Conflict Complexity is Lower Bounded by Block Sensitivity}
\author{Yaqiao Li\thanks{Universit{\'e} de Montr{\'e}al. \texttt{yaqiao.li@umontreal.ca}}}
\date{\today}

\maketitle

\begin{abstract}
We show conflict complexity of every total Boolean function, recently introduced in \cite{random2}  to prove a composition theorem of randomized decision tree complexity, is at least a half of its block sensitivity. We propose to compare conflict complexity with certificate complexity, and explain why it could be interesting.
\end{abstract}

\section{Introduction}  
Let $f: \{0,1\}^n \to \{0,1\}$ be a total Boolean function. The \emph{conflict complexity} of $f$, denoted by $\chi(f)$, is a complexity measure of $f$ that is recently defined in \cite{random2} and appears implicitly in \cite{random1}. Using this notion, \cite{random2} and \cite{random1} independently show an improved composition theorem for randomized decision tree complexity, a topic that is intensively studied in recent years, see \cite{anshu2017composition,ben2016randomized,random2, random1, ben2020tight}. Let $R(f)$ denote the randomized decision tree complexity of $f$ with a bounded error.  Let $g: \{0,1\}^m \to \{0,1\}$ be another Boolean function. The composed function $f\circ g: \{0,1\}^{nm} \to \{0,1\}$ is defined by $(f\circ g)(x_1, \ldots, x_{nm}) = f(g(x_1,\ldots, x_m), g(x_{m+1}, \ldots, x_{2m}), \ldots, g(x_{(n-1)m+1}, \ldots, x_{nm}))$. Priori to \cite{random2, random1}, it is shown in \cite{ben2016randomized} that $R(f\circ g) \ge \Omega\Big(R(f) \sqrt{R(g)/\log R(g)} \Big)$. Introducing the conflict complexity, it is proven in \cite{random2} that  $R(f\circ g) \ge \Omega(R(f) \chi(g))$ and $\chi(g) \ge \Omega(\sqrt{R(g)})$. Hence, $R(f\circ g) \ge \Omega(R(f) \sqrt{R(g)})$, removing the logarithmic factor in the previous lower bound.

Another important complexity measure of a Boolean function $f$ is its \emph{block sensitivity}, denoted by $\bs(f)$. Block sensitivity is firstly defined in \cite{bs}. Block sensitivity, its variants,  and the famous sensitivity conjecture (now a theorem by \cite{huang2019induced}) are widely studied and applied in complexity theory and combinatorics, see e.g. \cite{rubinstein1995sensitivity, survey, hatami2010variations, goos2014communication,ambainis2014tighter,kulkarni2016fractional} etc.
 The relation of block sensitivity  to other frequently used complexity measures of a Boolean function such as sensitivity, degree, decision tree complexity and certificate complexity etc, is relatively well understood. In particular, they are polynomially related, see the survey \cite{survey} and \cite{huang2019induced}. Explicit polynomial relations among different complexity measures are very useful. In particular, it allows one to transfer lower or upper bounds based on one complexity measure (e.g. degree) to another (e.g., block sensitivity). For example, \cite{huang2019induced} solves the sensitivity conjecture, which states  that block sensitivity is upper bounded by a polynomial of sensitivity, by showing that degree is upper bounded by a polynomial of sensitivity.

Let $D(f)$ denote the deterministic decision tree complexity of $f$, see definition in \cite{survey}. Later after we formally define $\chi(f)$, it will be clear that $\chi(f) \le D(f)$. Hence, by \cite{random2}, $\Omega(\sqrt{R(f)}) \le \chi(f) \le D(f)$. By \cite{survey},  $D(f)^{1/3} \le \bs(f) \le R(f)$. Hence, $\Omega(\sqrt{\bs(f)}) \le \chi(f) \le \bs(f)^3$. We improve this to a linear lower bound \footnote{Shortly after this work,  \cite{gavinsky2018composition} shows the \emph{max conflict complexity}, which is an upper bound of conflict complexity, is larger than the block sensitivity.} with an explicit coefficient.

\begin{theorem}  \label{thm:my}
For every non-constant total Boolean function $f: \{0,1\}^n \to \{0,1\}$, $\chi(f) \ge (\bs(f)+1)/2$.
\end{theorem}

The conflict complexity and block sensitivity are formally defined in Section~\ref{sec:def}. Theorem~\ref{thm:my} is proved in Section~\ref{sec:proof}. A question on comparing conflict complexity with certificate complexity and its possible consequences are discussed in Section \ref{sec:discussion}.

\section{Conflict Complexity and Block Sensitivity}  \label{sec:def}

Let $f: \{0,1\}^n \to \{0,1\}$ be a Boolean function. Let $\mu_0, \mu_1$ be two probabilisitic distributions on $f^{-1}(0)$ and $f^{-1}(1)$, respectively. Given $\cT$ as a deterministic decision tree\footnote{For more background on the definition of a decision tree and related complexity measures, see the survey \cite{survey}.}, which is a binary tree, that computes $f$ correctly. Let $v$ be a node of $\cT$, by an abuse of notation we also use $v$ to denote the Boolean variable that is queried at node $v$.  Every node $v \in \cT$  corresponds to a subcube of $\{0,1\}^n $ that is uniquely determined by the path leading from the root of $\cT$ to $v$. Denote
\begin{equation}  \label{eq:mu-v}
\mu^v_{0} = \mu_0|_v, \quad \mu^v_{1} = \mu_1|_v. 
\end{equation}
That is, they are the distributions of $\mu_0$ and $\mu_1$ conditioned on the subcube corresponding to $v$.  At node $v$, the decision tree $\cT$ branches to left or right according to $v = 0$ or $v = 1$, respectively. Denote
\begin{equation}  \label{eq:ab}
\alpha_v = \Pr_{x \sim \mu^v_0} [v = 0], \quad
\beta_v = \Pr_{x \sim \mu^v_1} [v = 0],
\end{equation}
where $x \sim \mu^v_0$ (resp. $x \sim \mu^v_1$) means to sample $x \in \{0,1\}^n$ according to the distribution $\mu^v_0$ (resp. $\mu^v_1$).

Consider the following random walk on the tree $\cT$ as follows: at node $v$, it goes to its left child (where $v=0$) with probability $\min\{\alpha_v, \beta_v\}$; it goes to its right child (where $v=1$)  with probability $1 - \max\{\alpha_v, \beta_v\}$; it stops at $v$ with probability $|\alpha_v - \beta_v|$.  We call a node $v \in \cT$ a \emph{last query node} if after the value of $v$ is queried, the tree outputs accordingly. Alternatively, the two nodes after a last query node are two leaves of the tree $\cT$. It is easy to see that $|\alpha_v - \beta_v| = 1$ if $v$ is a last query node. In particular, the random walk always stops once it reaches a last query node. Intuitively, conflict complexity denotes the expected number of nodes the random walk has visited when it stops. Note that we count the node at which the random walk stops.

\begin{definition}[\cite{random2}]  \label{def:conflictcomplexity}
Let $\mu_0, \mu_1, \cT$ be as defined before. Let $X = X(\mu_0, \mu_1, \cT)$ be the random variable taking values in $\N = \{1,2, \ldots, n\}$ that represents the number of nodes the random walk has visited when it stops. The \emph{conflict complexity} of $f$, denoted by $\chi(f)$, is defined as
\begin{equation}  \label{eq:def-chi}
\chi(f) = \max_{\mu_0, \mu_1} \min_{\cT} \Ex X.
\end{equation}
\end{definition}

Since the random walk always stops if it reaches a last query node,  $\chi(f) \le D(f)$.

We proceed to define the block sensitivity.  Let $f: \{0,1\}^n \to \{0,1\}$, $x \in \{0,1\}^n$, and $B \subseteq \{1,2,\ldots,n\}$ be a subset. Denote $x^B \in \{0,1\}^n$ as the $n$-bit string obtained from $x$ by flipping all bits whose indices are in the subset $B$. A subset $B$ is said to be a \emph{sensitive block} for $x$ if $f(x) \neq f(x^B)$. Let $\bs(f,x)$ denote the maximal  number of disjoint sensitive blocks  of $x$. The \emph{block sensitivity} of $f$, denoted by $\bs(f)$, is defined as $\bs(f) = \max_{x\in \{0,1\}^n} \bs(f,x)$.

\section{Proof of Theorem~\ref{thm:my}}   \label{sec:proof}

\begin{proof}
We exhibit distributions $\mu_0$ and $\mu_1$, such that for every deterministic decision tree $\cT$ that computes $f$ correctly,  $\Ex X \ge (\bs(f)+1)/2$.   Let $k = \bs(f)$. One has $k \ge 1$ since $f$ is not a constant function. Let $z \in \{0,1\}^n$ be an input string that achieves the block sensitivity of $f$, and $B_1, \ldots, B_k \subseteq \{1,2,\ldots,n\}$ are the disjoint sensitive blocks of $z$. Denote $y_i = z^{B_i} \in \{0,1\}^n$ for $i=1, \ldots, k$. Without loss of generality assume $f(z) = 0$, then $f(y_1) = \cdots = f(y_k) = 1$. Let $\mu_0$ be the distribution that is supported on the single point $z$, and $\mu_1$ be the uniform distribution over $Y = \{y_1, \ldots, y_k\}$. That is, 
\begin{equation}   \label{eq:distributions}
\mu_0(z) = 1, \quad
\mu_1(y_i) = 1/k, \ i=1, \ldots, k.  
\end{equation}
Let $X$ be the random variable defined as in Definition \ref{def:conflictcomplexity}. Let $P$  denote the unique path that is travelled by the input $z$ in the decision tree $\cT$. Let $\ell$ denote the length (i.e., number of nodes) of $P$, obviously $\ell \ge k$ since at least one bit from each sensitive block of $z$ must be queried. Renaming variables if necessary, we may assume $x_1, \ldots, x_\ell$ are the successive nodes in the path $P$ where $x_1$ is the root and $x_\ell$ is the last query node. Note that $\mu_0^{x_1} = \mu_0$ and $\mu_1^{x_1} = \mu_1$.

We will analyze $\alpha_v$ and $\beta_v$, for $v=x_1, \ldots, x_\ell$. Consider $\alpha_v$ first. We claim that $\alpha_v = 1$ (resp.  $\alpha_v = 0$)  if  $v= 0$ (resp. $v= 1$)  in the path $P$. Indeed, since $P$ is the path travelled by $z$ and  $\mu_0(z) = 1$, for every node $v$ in the path $P$ it holds that $\mu^v_0(z) = 1$. Hence, the claim follows by definition \eqref{eq:ab}. 

We proceed to analyze $\beta_v$.  It is illuminating to firstly analyze $\beta_v$ at the root $v = x_1$. There are two cases according to whether $1 \in \cup_{j=1}^k B_j$. Note that here the number $1$ is the index of $x_1$. 
\begin{itemize}
\item[(i)] $1 \not\in \cup_{j=1}^k B_j$. This implies $y_{i,1}= z_1$ for all $y_i \in Y$. Hence, $\beta_{x_1} = \alpha_{x_1}$, showing that   
$\Pr[X=1] = | \alpha_{x_1} - \beta_{x_1}| = 0$, $\Pr[\text{the random walk reaches}\ x_2] = 1$, and $\mu^{x_2}_1 = \mu_1$.

\item[(ii)] $1 \in B_j$ for some $j \in \{1, \ldots, k\}$. Since $B_j \cap B_i =\emptyset$ for every $i\neq j$, there is exactly one such $B_j$. Without loss of generality, we assume $z_1 = 0$. Hence $\alpha_{x_1} = 1$. Since $z_1 = 0$, one has $y_{j,1} = 1$ and $y_{i,1} = 0$ for all other $i \neq j$. Hence, 
\[
\beta_{x_1} = \Pr_{x \sim \mu_1} [x_1 = 0] = \Pr_{y_i \sim Y}[y_{i,1} = 0] = (k-1)/k.
\]
Therefore,
\[
\Pr[X=1] = |\alpha_{x_1} - \beta_{x_1}| = 1/k.
\]
Besides, $\Pr[\text{the random walk reaches}\ x_2] = (k-1)/k$, and $\mu^{x_2}_1$ is a uniform distribution over $Y - y_j$, a set of $k-1$ elements. 
\end{itemize}

We claim that the phenomenon discussed in (i) and (ii) is true in general: for every $r = 1, \ldots, \ell$, either $\Pr[X=r] = 0$ or $\Pr[X=r] = 1/k$.   Since $\ell \ge k$, this immediately implies the desired bound
\[
\Ex X \ge \sum_{i=1}^k i \cdot (1/k) = (k+1)/2.
\]

We proceed to show the claim. Consider an arbitrary node $v = x_r$ in the path $P$. Let 
\[
\cA^v = \{j: B_j \cap \{1, \ldots, r-1\} = \emptyset\}, \quad Y^v = \cup_{j \in \cA^v} \{y_j\}.
\]
Intuitively, $Y^v$ is the set of $y_j$ that are still ``active'' at node $v$ (i.e., not deviated from the path $P$ before $v$). By a similar analysis as we did for the root, we know that $\mu^v_1$ is a uniform distribution on $Y^v$. Hence, at node $v=x_r$, the random walk stops with probability, 
\begin{equation}  \label{eq:prob-stop}
|\alpha_v - \beta_v| = 
\begin{cases}
0, &\quad r \not\in \cup_{j \in \cA^v} B_j; \\
1/ |\cA^v|, &\quad r \in \cup_{j \in \cA^v} B_j.
\end{cases}
\end{equation} 
On the other hand, for any two successive nodes $x_i, x_{i+1}$ in $P$, 
\[
\Pr[\text{the random walk branches from\ } x_i \text{ to } x_{i+1}]
= 
\begin{cases}
1, &\quad i \not\in \cup_{j \in \cA^{x_i}} B_j; \\
( |\cA^{x_i}| - 1)/ |\cA^{x_i}|, &\quad i \in \cup_{j \in \cA^{x_i}} B_j.
\end{cases}
\]
This implies 
\begin{equation} \label{eq:prob-reach}
\Pr[\text{the random walk reaches $v$}] = |\cA^v| / k. 
\end{equation}
Apply  \eqref{eq:prob-stop} and \eqref{eq:prob-reach}, we get
$\Pr[X=r] = \Pr[\text{the random walk reaches $v$}] \cdot |\alpha_v - \beta_v|$,
is either $0$ or $1/k$ as claimed.
\end{proof}

\section{Discussion} \label{sec:discussion}
It seems an interesting problem to compare conflict complexity with certificate complexity, as explained below. Given $B\subseteq \{1,2,\ldots,n\}$ and $x \in \{0,1\}^n$. Let $x|_B$ denote the $|B|$-bit substring obtained by restricting  $x$ to the indices in $B$.  For every Boolean function $f: \{0,1\}^n \to \{0,1\}$, let $C_x(f)$ denote the minimal integer $0\le k \le n$ such that there exists a subset $B\subseteq \{1,2,\ldots,n\}$ with $|B|=k$ satisfying $f(y)= f(x)$ whenever $y|_B = x|_B$.  The \emph{certificate complexity} of $f$, denoted by $C(f)$, is defined as $C(f) = \max_{x\in \{0,1\}^n} C_x(f)$. The following holds,  
\[
(\sqrt{C(f)} +1)/2 \le \chi(f) \le C(f)^2.
\]
Indeed, $\sqrt{C(f)} \le \bs(f) \le D(f) \le C(f)^2$ where $D(f)$ is the deterministic decision tree complexity of $f$, see \cite{survey}. Hence,  the inequality follows from Theorem \ref{thm:my} and the fact that  $\chi(f) \le D(f)$.

\noindent {\bf Question}: $\chi(f) \ge \Omega(C(f))$ or $\chi(f) \le O(C(f))$\ ? 

The answer could be neither. However, a positive answer  in either case has interesting consequences. 

\begin{itemize}
\item The case $\chi(f) \ge \Omega(C(f))$. Let $R(f)$ denote the randomized decision tree complexity of $f$ with a bounded error (see definition in \cite{survey}). Via $C(f) \ge \sqrt{D(f)}$ (see \cite{survey}), by \cite{random2} one has $R(f \circ g) \ge \Omega\Big(R(f) \chi(g) \Big) \ge \Omega\Big(R(f)  \sqrt{D(g)}\Big)$, improving the lower bound $R(f\circ g) \ge \Omega\Big(R(f) \sqrt{R(g)}\Big)$ in \cite{random2, random1, gavinsky2018composition}.

\item The case $\chi(f) \le  O(C(f))$. This has two consequences. 

\begin{itemize}
\item[(1)] Consider the $\AND$ of $\OR$ tree $\AND_n \circ \OR_n: \{0,1\}^{n^2} \to \{0,1\}$, which is the composition of $n$-bit $\AND$ function with $n$-bit $\OR$ function. It is easy to see that $C(\AND_n \circ \OR_n) = n$ and $\bs(\AND_n) = \bs(\OR_n) = n$, see \cite{survey}. Hence, Theorem \ref{thm:my} implies $\chi(\AND_n) \chi(\OR_n) \ge \Omega(n^2)$, but $\chi(\AND_n \circ \OR_n) \le O\Big( C(\AND_n \circ \OR_n) \Big)   = O(n)$. This implies $\chi(\AND_n \circ \OR_n) \le O\Big(\sqrt{\chi(\AND_n)  \chi(\OR_n)}\Big)$. In particular, it shows that $\chi(f\circ g) = \Theta\Big(\chi(f) \chi(g)\Big)$ can not hold.

\item[(2)] In \cite{jain2010partition} it is shown  $R(\AND_n \circ \OR_n) = \Theta(n^2)$. Hence, $\chi(\AND_n \circ \OR_n) \le O\Big(\sqrt{R(\AND_n \circ \OR_n)}\Big)$. This implies the lower bound $\chi(f) \ge \Omega(\sqrt{R(f)})$ in \cite{random2,gavinsky2018composition} is tight.  
\end{itemize}
\end{itemize}

The same question can be asked for the \emph{max conflict complexity} as introduced in \cite{gavinsky2018composition}.

\section*{Acknowledgement}

The author thanks the host of Simons Institute for the Theory of Computing where this work was done.

\bibliography{mybib}{}

\begin{thebibliography}{10}

\bibitem{ambainis2014tighter}
Andris Ambainis, Mohammad Bavarian, Yihan Gao, Jieming Mao, Xiaoming Sun, and
  Song Zuo.
\newblock Tighter relations between sensitivity and other complexity measures.
\newblock In {\em International Colloquium on Automata, Languages, and
  Programming}, pages 101--113. Springer, 2014.

\bibitem{anshu2017composition}
Anurag Anshu, Dmitry Gavinsky, Rahul Jain, Srijita Kundu, Troy Lee, Priyanka
  Mukhopadhyay, Miklos Santha, and Swagato Sanyal.
\newblock A composition theorem for randomized query complexity.
\newblock In {\em 37th IARCS Annual Conference on Foundations of Software
  Technology and Theoretical Computer Science}, page~1, 2018.

\bibitem{ben2020tight}
Shalev Ben-David and Eric Blais.
\newblock A tight composition theorem for the randomized query complexity of
  partial functions.
\newblock {\em arXiv preprint arXiv:2002.10809}, 2020.

\bibitem{ben2016randomized}
Shalev Ben-David and Robin Kothari.
\newblock Randomized query complexity of sabotaged and composed functions.
\newblock {\em Theory of Computing}, 14(1):1--27, 2018.

\bibitem{survey}
Harry Buhrman and Ronald De~Wolf.
\newblock Complexity measures and decision tree complexity: a survey.
\newblock {\em Theoretical Computer Science}, 288(1):21--43, 2002.

\bibitem{random1}
Dmitry Gavinsky, Troy Lee, and Miklos Santha.
\newblock On the randomised query complexity of composition.
\newblock {\em arXiv preprint arXiv:1801.02226}, 2018.

\bibitem{gavinsky2018composition}
Dmitry Gavinsky, Troy Lee, Miklos Santha, and Swagato Sanyal.
\newblock A composition theorem for randomized query complexity via max
  conflict complexity.
\newblock In {\em 46th International Colloquium on Automata, Languages, and
  Programming (ICALP 2019)}, volume 132, page~64. Schloss
  Dagstuhl--Leibniz-Zentrum fuer Informatik, 2019.

\bibitem{goos2014communication}
Mika G{\"o}{\"o}s and Toniann Pitassi.
\newblock Communication lower bounds via critical block sensitivity.
\newblock In {\em Proceedings of the forty-sixth annual ACM symposium on Theory
  of computing}, pages 847--856. ACM, 2014.

\bibitem{hatami2010variations}
Pooya Hatami, Raghav Kulkarni, and Denis Pankratov.
\newblock Variations on the sensitivity conjecture.
\newblock {\em Theory of Computing}, pages 1--27, 2011.

\bibitem{huang2019induced}
Hao Huang.
\newblock Induced subgraphs of hypercubes and a proof of the sensitivity
  conjecture.
\newblock {\em Annals of Mathematics}, 190(3):949--955, 2019.

\bibitem{jain2010partition}
Rahul Jain and Hartmut Klauck.
\newblock The partition bound for classical communication complexity and query
  complexity.
\newblock In {\em 2010 IEEE 25th Annual Conference on Computational
  Complexity}, pages 247--258. IEEE, 2010.

\bibitem{kulkarni2016fractional}
Raghav Kulkarni and Avishay Tal.
\newblock On fractional block sensitivity.
\newblock {\em Chicago J. Theor. Comput. Sci}, 2016:2, 2016.

\bibitem{bs}
Noam Nisan.
\newblock Crew prams and decision trees.
\newblock {\em SIAM Journal on Computing}, 20(6):999--1007, 1991.

\bibitem{rubinstein1995sensitivity}
David Rubinstein.
\newblock Sensitivity vs. block sensitivity of boolean functions.
\newblock {\em Combinatorica}, 15(2):297--299, 1995.

\bibitem{random2}
Swagato Sanyal.
\newblock A composition theorem via conflict complexity.
\newblock {\em arXiv preprint arXiv:1801.03285}, 2018.

\end{thebibliography}
\bibliographystyle{plain}

\end{document}